\documentclass[aps,floatfix,showpacs,preprint,superscriptaddress]{revtex4}
\usepackage{mathrsfs}

\usepackage{graphicx}
\usepackage{dcolumn}
\usepackage{amsmath}
\usepackage{longtable}

\begin{document}

\title{Relation Between Chiral Susceptibility and Solutions of Gap Equation in Nambu--Jona-Lasinio Model}
\author{Yue Zhao}
\affiliation{Department of Physics, Peking University, Beijing
100871, China}

\author{Lei Chang}
\affiliation{Department of Physics, Peking University, Beijing
100871, China}

\author{Wei Yuan}
\affiliation{Department of Physics, Peking University, Beijing
100871, China}

\author{Yu-xin Liu}
\email[corresponding author, ]{yxliu@pku.edu.cn}
\affiliation{Department of Physics, Peking University, Beijing
100871, China} \affiliation{The Key Laboratory of Heavy Ion Physics,
Ministry of Education,Beijing 100871, China } \affiliation{Center of
Theoretical Nuclear Physics, National Laboratory of Heavy Ion
Accelerator, Lanzhou 730000, China}
%

\date{\today}

\begin{abstract}
We study the solutions of the gap equation, the thermodynamic
potential and the chiral susceptibility in and beyond the chiral
limit at finite chemical potential in the Nambu--Jona-Lasinio (NJL)
model. We give an explicit relation between the chiral
susceptibility and the thermodynamic potential in the NJL model. We
find that the chiral susceptibility is a quantity being able to
represent the furcation of the solutions of the gap equation and the
concavo-convexity of the thermodynamic potential in NJL model. It
indicates that the chiral susceptibility can identify the stable
state and the possibility of the chiral phase transition in NJL
model.
%
\end{abstract}


\maketitle

\section{Introduction}
Quantum chromodynamics(QCD) is a non-Abelian gauge theory. The proof
of its renormalizability~\cite{hooft} and discovery of ultraviolet
asymptotic freedom~\cite{Gross73, Polit73} have been milestones in
its acceptance as the theory of the strong interaction. For large
momentum, the coupling becomes very weak, then perturbation theory
is appropriate to carry out the calculations. However, for small
momentum, the coupling grows quite strong and adequate methods have
to be implemented to study the nonperturbative phenomena, such as
confinement, dynamical chiral symmetry breaking (DCSB) and bound
state formation. Among these characters DCSB is fundamentally
important. For example, it is responsible for the generation of
large constituent-like masses for dressed-quarks in QCD and also is
the keystone in the realization of Goldstone's theorem through
pseudoscalar mesons in QCD. It is generally believed that, at
sufficiently high temperature and/or density, the QCD vacuum
undergoes a phase transition into a chiral symmetric phase. This
chiral phase transition plays an essential role in studying the
structure of some astro-objects and the evolution of the early
universe, which may be experimentally realized in ultra high energy
heavy-ion collisions. At finite temperature, the lattice simulation
is powerful for studying the chiral phase transition. It is now
under developing also for finite chemical potential. However,
effective theories of QCD are still necessary, even powerful, for
various nonperturbative phenomena including the phase transition.
There have been many approaches and models exhibiting such a
character, such as Dyson-Schwinger equation (DSE)
approach~\cite{Craig01,Alkofer01,Craig02,Watson01,Harada99,Craig03,Alkofer02,Alkofer03,Craig04,Pennington04,Alkofer04,YCL06},
Nambu--Jona-Lasinio (NJL)
model\cite{NJL61,Volk84,Hat84,Klevansky,Alkofer96}, chiral quark
model\cite{Faessler,CQSM1,CQSM2,Glozman,ZhangZY}, Global Color
Symmetry Model (GCM)~\cite{CR858,FT9127,Tan97,LLZZ98}, quark-meson
coupling model\cite{Guichon}, quark mean field model\cite{Toki}, and
so on.

In these QCD-like theories, one usually tries to find at first the
solutions of the equations satisfied by order parameters of the
phase transition, and then to determine which solution is stable at
certain condition by analyzing the thermodynamic potential. In
Ref.~\cite{YCL06}, based on the DSE approach, some of us have shown
that the chiral susceptibility is a quantity which could describe
the dependence of the chiral condensates or dressed quark mass on
the chemical potential and the current quark mass at the first order
approximation. The chiral susceptibility $\chi$ is usually defined
as the first order response of the order parameter or of the dressed
quark mass with respect to the current quark mass. Furthermore
Ref.~\cite{YCL06} suggests that the chiral susceptibility could be
used to represent the possibility of the chiral and other phase
transitions. However, the explicit relation between the chiral
susceptibility and the solutions of the gap equation has not yet
been given. Then we study the property of $\chi$ in this paper, and
discuss that it is just the quantity to characterize the properties
of the furcations of the gap-equation's solutions.

The DSE approach and the NJL model both have its quark (gap)
equation, and could get into the same form under some
approximation. Comparing these two approaches, we could see that
DSE is a superior and accurate model because the NJL model takes
the point-like interaction approximation for the gluon-mediated
interaction among quarks. However, finding the solutions of DSEs
and determining which solution is the physical solution remain to
be a difficult problem, because as the higher order loops are
taken into account, more equations need to be solved, and more
complicated even impossible to give the explicit expression of
thermodynamic potential. However there are not so many
difficulties in the NJL model. Thus if we could find some
quantities which may characterize the properties of the solutions
in a simple model, for example the NJL model, and generalize it to
more advanced models such as DSE, it will be much helpful in
studying the property of DCSB. We then study the solutions of the
gap equation and the thermodynamic potential in the NJL model and
discuss their relation with the chiral susceptibility in this
paper.

The paper is organized as follows. In Section II, we describe
briefly the main points of the NJL model and the corresponding gap
equation for quark. In Section III, we discuss the solutions of the
gap equation and the chiral susceptibility. In Section IV, we
represent the feature of the thermodynamic potential in the NJL
model and discuss the phase transitions. In Section V, we give the
relation between the chiral susceptibility and the thermodynamic
potential in the NJL model. Finally in Section VI, we give a summary
and brief remarks.

\section{Brief Description of the NJL Model}

The Nambu--Jona-Lasinio (NJL) model was originally developed as a
theory to study the interaction of nucleons through an effective
two-body interaction~\cite{NJL61}. Then it was extended to describe
the interaction in the quark degrees of freedom. Because of its
simplicity, the NJL model is useful for us to understand the process
of chiral symmetry breaking. Also it could explicitly show us the
Goldstone modes.

As an effective approach, NJL model has been used to study the
chiral phase transition in the matter at finite temperature and
baryon chemical potential, the color superconductivity at moderate
baryon density, the properties of some asymmetry matter and the
structure of some astro-objects (see for example
Refs.~\cite{MHuang,Baldo,Blaschke,Bentz,Ruster,Barducci,Buballa05,
Aguilera,He,Castorina,Forbes,Shao,Ghosh,Ohsaku,Hands,QWang}). In
this paper we just discuss the chiral dynamical behavior and start
with the simple Lagrangian of the NJL model
\begin{equation}
\mathscr{L}_{NJL}=\bar{q}(i \gamma^\mu \partial_\mu-m)q +
G[(\bar{q}q)^2+(\bar{q} i \gamma_5 \vec{\tau}q)^2] \, ,
\end{equation}
where $m$ is the current quark mass, $\vec{\tau}$ are Pauli
matrices and $G$ is the coupling constant.

Under the Hartree approximation (i.e., the random phase
approximation (RPA)), the interaction terms could be substituted as
\begin{displaymath}
(\bar{q}O q)^2 \longrightarrow 2\langle\bar{q}Oq\rangle \bar{q}O q
- \langle\bar{q}Oq\rangle^2 \, ,
\end{displaymath}
in which $O$ could be the interaction matrix $1$, $\gamma_5$,
$\gamma_\mu$, $\gamma_5 \gamma_\mu$. If we only take the scalar
condensation into account, the Lagrangian could be written as
\begin{equation}
\mathcal{L}_{NJL}=\bar{q} (i \gamma^\mu \partial_\mu - m)q +
2G\langle\bar{q}q\rangle \bar{q}q - \frac{(M-m)^{2}}{4G}
       =\bar{q}(i \gamma^\mu \partial_\mu - M)q - \frac{(M-m)^{2}}{4G} \, ,
\end{equation}
where $M=m-2G \langle \bar{q}q\rangle$ is the constituent quark
mass. Under the Hartree approximation, the self energy of a quark
is generated by the local four-fermion interaction. After some
derivation, we can write the $M$ as
\begin{eqnarray}
 M&=&m + 2iG\int \frac{d^4p}{(2\pi)^4} Tr S(p)\nonumber\\
  &=&m + 8 N_f N_c i G\int \frac{d^4p}{(2\pi)^4} \frac{M}{p^2-M^2}
  \, ,
\end{eqnarray}

Using the standard technique of thermal field theory, we could
directly calculate the contribution of the quark loops and
polarization diagrams at finite temperature $T$ and finite
chemical potential $\mu$. Thus, we could get the gap equation with
the variables of $T$ and $\mu$ as
\begin{equation}
\frac{M-m}{G}-4 N_c N_f \int \frac{d^3p}{(2\pi)^3}
\frac{M}{E_p}(1-n_p(T,\mu)-\bar{n_p}(T,\mu))=0\, ,
\end{equation}
where $E_p=\sqrt{p^2+M^2}$, $n_p(T,\mu)$ and $\bar{n_p}(T,\mu)$
are Fermi occupation number of quarks, anti-quarks, respectively,
with
\begin{displaymath}
n_p(T,\mu)=\frac{1}{e^{(E_p-\mu)/T}+1} \, ,
\end{displaymath}
\begin{equation}
\bar{n_p}(T,\mu)=\frac{1}{e^{(E_p+\mu)/T}+1} \, .
\end{equation}
If we set $T=0$, $\mu=0$, Eq.~(4) will share the same form as
Eq.~(3) after the integration over $p_0$.

\section{Solutions of the Gap Equation and Chiral Susceptibility}

%

We constrain ourself at present with zero temperature and finite
chemical potential. After some derivation from Eqs.~(4), the gap
equations could be written, respectively, as
\begin{eqnarray}
\frac{M-m}{G}&=&8 N_c \int^\Lambda_0 \frac{p^2
dp}{2\pi^2}\frac{M}{E_p} \, , \qquad \quad \mbox{for} \; \mu < M \,
,\\ \frac{M-m}{G}&=&8 N_c \int^\Lambda_{k_f} \frac{p^2
dp}{2\pi^2}\frac{M}{E_p} \, , \qquad \quad \mbox{for} \; \mu > M \,
,
\end{eqnarray}
where $m$ is the current quark mass, $\mu=\sqrt{M^2+k_f^2}$ with
$k_f$ being the Fermion momentum, and $\Lambda$ is the cut-off of
the three-momentum.

From the gap equation in Eq.~(7), we can easily obtain the first
order derivative of the constituent quark mass with respect to the
current quark mass (we discuss only the states with $\mu>M$ since
those with $\mu < M$ are constant. In addition, it is important to
mention that we do not take the chiral limit of the gap equation
before we carry out the derivative with respect to the $m$, and then
we set $m=0$ to get the result in the chiral limit). The obtained
chiral susceptibility can be written as
\begin{eqnarray}\label{chi}
\chi &=& \frac{\partial M}{\partial m}  \nonumber \\
     &=&\frac{1}{1-8 N_c G
\frac{\partial^2}{\partial M^2} \int^\Lambda_{k_f} \frac{p^2
dp}{2\pi^2}E_p-8 N_c \mu G \frac{\partial^2}{\partial M^2}
\int^\Lambda_{k_f} \frac{p^2 d p}{2\pi^2}} \, .
\end{eqnarray}

In the following we should discuss the solutions of gap equation and
investigate the relation between the solutions and chiral
susceptibility in the chiral limit and beyond the chiral limit,
respectively.

\subsection{In the Chiral Limit}

We concentrate ourself at present in the chiral limit, i.e.,
setting $m=0$ in Eqs. (6) and (7). To solve these gap equations,
we take the parameters as $\Lambda=587.90$~MeV and $G \Lambda^2 =
2.44$, with which the experimental data of the mass and the decay
constant of pion are reproduced well~\cite{Buballa05}. The
obtained solutions of the gap equations are illustrated in Fig.~1.

\begin{figure}[!ht]
\includegraphics[scale=1.0,clip]{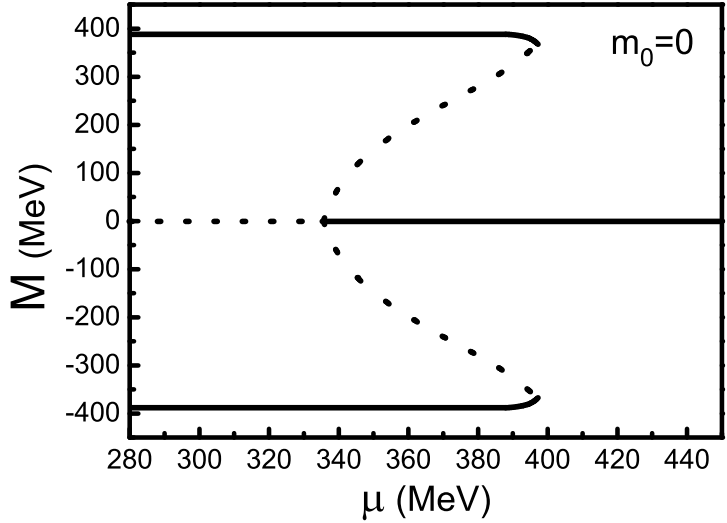}
\caption{Solutions of the gap equation in terms of the chemical
potential at the chiral limit and zero temperature (with
parameters being set as $\Lambda=587.9 MeV$ and $G \Lambda^2 =
2.44$).}
\end{figure}

The figure shows evidently that, at zero chemical potential, there
exist three solutions of the gap equation, $M=\pm 387.92$~MeV and
$M=0$~MeV. The solutions $M=\pm 387.92$~MeV are usually referred
to as the positive, negative Nambu solution, respectively. And
$M=0$~MeV is denoted as the Wigner solution. Wigner solution
corresponds to the chiral symmetric vacuum, and the Nambu
solutions mean that the vacuum has already been broken into a
chiral asymmetric state, and the condensate induces the dress of
quark mass. With the increasing of chemical potential the Nambu
solutions and Wigner solution would take on different behaviors.

We firstly discuss the case with chemical potential $\mu<|M|$. From
Eq.~(6), we know easily that the equation is independent of chemical
potential $\mu$. Thus, the Nambu solutions are constants while we
change the value of $\mu$ and they would only exist when $0 \le \mu
< 387.92$~MeV.

From Eq.~(7) we can infer that, when $\mu>|M|$, the solutions of gap
equation are much more complicated. From Fig.~1, we could see that,
when $0 \le \mu < 335.59$~MeV, there exists only a Wigner solution
with $M=0$~MeV. At the point $(335.59,0)$, there emerges a
trifurcation, one of which remains zero, another two becomes
positive, negative, respectively. This indicates that the Wigner
solution splits at the point $\mu = 335.59$~MeV. As we increases the
chemical potential continuously, The zero solution maintains zero
and the non-zero solutions changes with the absolute value
increased. However there are nothing special until $\mu=
387.92$~MeV. We have known that, from this point, the solutions of
Eqs.~(6) and (7) break the constraint $\mu<|M|$, thus there are no
longer constant solutions beyond this chemical potential. Moreover,
just from such a $\mu$, two more solutions of Eq.~(7) appear, and
these two solutions are accurately generated from the points at
which the constant Nambu solutions end. At first glance, it seems to
be surprising for us to separate the chemical potential to two parts
with $\mu < M$, $\mu >M$, respectively. Nevertheless, we notice that
the chemical potential is originally involved in the exponential
function in the denominator of the occupation number, as we take the
limit of zero temperature, the occupation number function separates
into the two parts correspondingly. It induces then the gap equation
to two parts. If we take finite temperature, such a separation does
not emerge. Then there will not be any problems for the connection
between the solution with $\mu < M$ and that with $\mu > M$. In the
general situation, the physical content should vary gradually if we
change a physical variable slowly. So it is quite natural for the
appearance of junctions at $\mu = 387.92$~MeV. When $387.92< \mu <
397.23$~MeV, there are totally five solutions for Eq.~(7), two of
them are positive, one of them is zero and another two are negative.
As $\mu=397.23$~MeV, the two positive solutions coalesce together.
It manifests that a bifurcation exists around $\mu \leq 397.23$~MeV.
And the negative solutions have the same behavior as the positive
ones.

In general, furcation means that more than one states are possibly
to appear. Then the bifurcation or trifurcation of the solutions
of the gap equation in NJL model is quite important in
understanding the possible states in the NJL model. If one could
find out the positions of the furcations, it would be easy to
obtain the evolution of the solutions and the appearance of the
possible states with respect to the chemical potential. One can in
turn characterize the property of the stable states easily.

Our purpose is to find out a quantity which could represent the
variation behavior of the solutions. In the above description, we
have shown that both of the Wigner solution and the Nambu solution
of the gap equation in the NJL model involve furcations. Then if a
quantity could characterize the appearance of the furcations, it
could be taken as a signature to study the feature of the
solutions of the gap equation. Furthermore, because we should not
constrain ourselves in the NJL model, it should be a quantity
easily to write out, and could be generalize to other formalism,
such as the DSE approach.

With the same parameters as taken to get Fig.~1, we calculate the
chemical potential dependence of the chiral susceptibility. The
obtained result is illustrated in Fig.~2.
\begin{figure}[!ht]
\includegraphics[scale=1.40,clip]{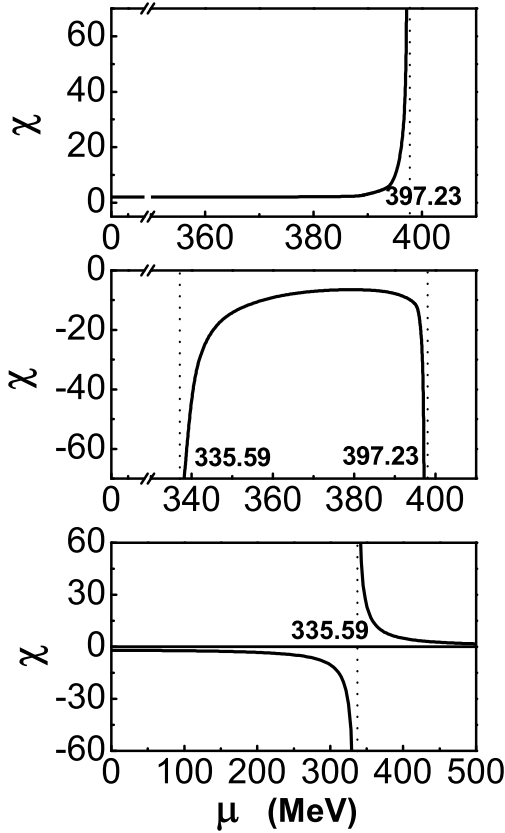}
\caption{Calculated chemical potential dependence of the chiral
susceptibility (with the parameters being taken as the same as for
Fig.~1), upper panel for the Nambu solutions, middle panel for the
nonzero Wigner solutions and the lower panel for the zero Wigner
solution.}
\end{figure}
The figure shows evidently that every appearance of the furcation in
of the solutions ($\mu = 335.59$~MeV, $397.23$~MeV) corresponds to
the emergence of a singularity of the chiral susceptibility. Because
the definition of $\chi$ is based on the gap equation, only the
constituent quark mass which satisfies the gap equation could be
substituted into Eq.~(8). Thus, as we combine Eq.~(7) and the
equation for the chiral susceptibility $\chi$ to be divergent, we
could gain all the furcations of the solutions. We could find that
this means of finding the furcations could be generalized to other
formalism, such as the DSE approach. In DSE approach, the quark
equation corresponds to the gap equation under some approximations
of QCD. We could also give the definition of chiral susceptibility
on the basis of the quark equation. So it could be used to study the
properties of solutions in DSE approach as well.

\subsection{Beyond the Chiral Limit}

In last subsection, we discuss the solutions of the gap equation
of quark and the chiral susceptibility in the chiral limit in NJL
model. Now we go beyond the chiral limit. With the interaction
parameters being taken as the same as those in the chiral limit
and the current quark mass being set as $m=5.6$~MeV, which is
consistent with the conventional choice, we solve Eq.~(6) and
Eq.~(7). The obtained results are illustrated in Fig.~3.

\begin{figure}[!ht]
\includegraphics[scale=0.70,clip]{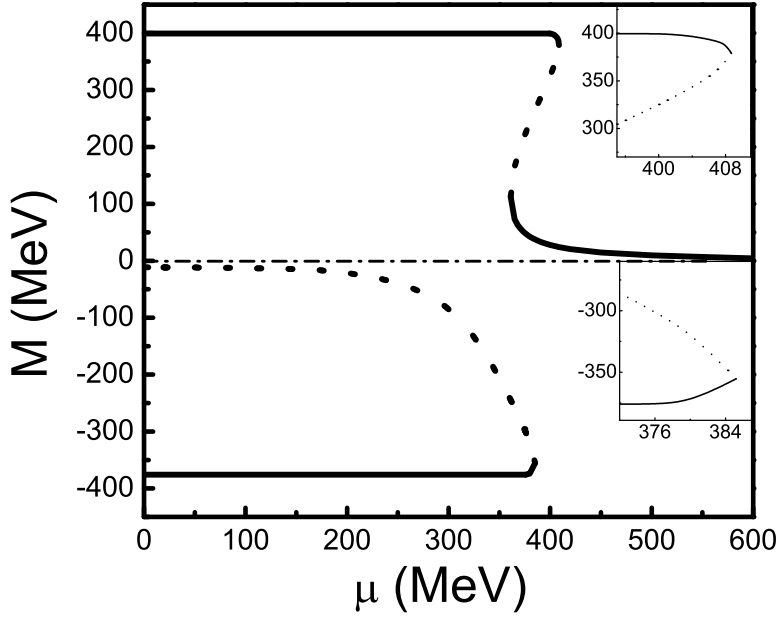}
\caption{Solutions of the gap equation beyond the chiral limit and
at zero temperature as a function of the chemical potential (with
the same parameters as those for Fig.~1 and a current quark mass
$m=5.6$ MeV).}
\end{figure}

At zero chemical potential there exist three different solutions
in such a model which has been carefully discussed in
Ref.~\cite{LChang} and implemented to study the relation of
explicit chiral symmetry breaking and DCSB. In this paper we
extend this case to finite chemical potential and study the
chemical potential dependence of these solution. It is evidently
that, in the case of $\mu<M$, since the gap equation is still
independent of the chemical potential $\mu$, it would give
constant solutions with respect to the chemical potential and
these solutions would disappear at the $\mu$ which is equal to the
absolute value of the $M$, respectively. In the case of $\mu > M$,
Eq.~(7) is obviously depending on the chemical potential $\mu$,
because of the appearance of the Fermi momentum $k_{f}$. It should
then be solved numerically. The obtained numerical results of the
solutions against the chemical potential $\mu$ are illustrated in
Fig.~3.

The figure shows apparently that, at $\mu =0$~MeV, Eq.~(6) has three
solutions $M = 399.44$~MeV, $-11.64$~MeV and $-375.89$~MeV. At the
point $(\mu , M)= (-11.64,11.64)$, the solution $M=-11.64$~MeV of
Eq.~(6) disappears and a solution of Eq.~(7) starts accurately from
this point. As the chemical potential $\mu$ increases, the absolute
value of this solution increases. In the region $\mu \in (11.64,
361.08)$~MeV, there is only such a solution for Eq.~(7). As $\mu$
takes a value $361.08$~MeV, another solution with $M=112.39$~MeV
emerges. Furthermore, from such a point, a bifurcation of the
solutions appears, in which one raises with the increasing of $\mu$
and the other decreases. When $\mu$ reaches $375.89$~MeV, the
solution $M=-375.89$~MeV for Eq.~(6) ends and a solution with the
same value appears for Eq.~(7). With the increasing of chemical
potential, the value of such a negative solution of Eq.~(7)
increases. As the chemical potential reaches $\mu = 385.20$~MeV, the
two negative solutions coalesce at $(\mu , M) =
(385.20,-355.37)$~MeV. Then the negative solutions disappear, and
remain a bifurcation bellow that point. For the positive solution,
the point $(\mu , M ) = (399.44, 399.44)$ is the end of the constant
solution of Eq.~(6), as well as the starting of another solution of
Eq.~(7). With further increase of the chemical potential, this
solution joints with the upper solution generated at the point $(\mu
, M) = (361.08, 112.39)$ as the chemical potential $\mu =
408.70$~MeV. From this point, there is only one solution exists for
Eqs.~(6) and (7) (in fact, only for Eq.~(7)), and it gradually
approaches to zero when chemical potential increases to positive
infinite.

As defined above, we could get the explicit form of chiral
susceptibility from the gap equation, and it has the same form as
Eq.~(8). If we require $\chi$ divergent and combine such a
requirement with Eq.~(8), we could get three solutions for the
chemical potential $\mu$. The obtained results are exactly the
points for the bifurcations of the solutions to appear. The detailed
numerical results of the $\chi$ in terms of the $\mu$ is displayed
in Fig.~4.
\begin{figure}[!ht]
\includegraphics[scale=1.4,clip]{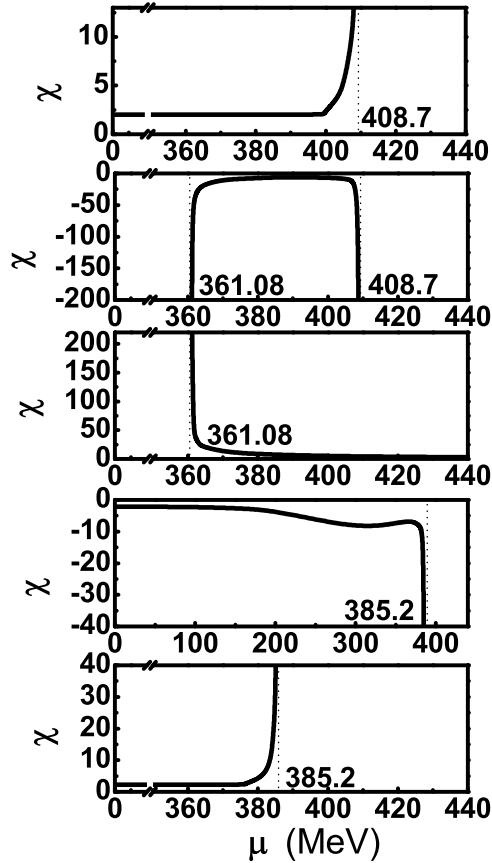}
\caption{Calculated chemical potential dependence of the chiral
susceptibility beyond the chiral limit (with the parameters being
taken as the same as for Fig.~3). The panels from the top to the
bottom corresponds to the solution illustrated in Fig.~3 from the
uppermost to the lowermost sequently.}
\end{figure}
Once again, we show that the singularity of chiral susceptibility
corresponds to the furcation of the solutions, and this property
is not constrained in the chiral limit. We could then take the
chiral susceptibility $\chi$ to characterize the critical points,
at which the properties of the gap equation's solutions changes.
Therefore the chiral susceptibility would be useful in picking out
the physical states of the system.

In addition, it may also be interesting to discuss how the
solutions beyond the chiral limit evolve from those in the chiral
limit. In Ref.~\cite{LChang}, some of us and collaborators show
that, there exists a convergent radius for the current quark mass,
within which the dynamical (constituent) mass function can be
expanded as a series in terms of the current mass. Then, if the
current quark mass is small, the $m \chi$ could be a quite good
approximation for the contribution of the current mass effect.
Since the chiral susceptibility $\chi$ is positive for the Nambu
solutions, and negative for the Wigner solutions, if the chemical
potential is less than a corresponding critical value, we can then
understand that the positive, negative Nambu solution beyond the
chiral limit comes from the positive, negative Nambu solution with
an increase of value, respectively. And the Wigner solutions
beyond chiral limit arises from the Wigner solutions with a
decrease of value, they separate into two distinct branches due to
the divergence of the chiral susceptibility.

\section{Analysis of Chiral Phase Transition }

Solving the gap equation could only get the possible physical states
of a system. To obtain the stable state which holds the lowest
energy, one should compare the thermodynamic potentials
corresponding to the solutions. In NJL model, it is not difficult to
obtain the thermodynamic potential $\Omega$ from the Lagrangian by
applying the standard technique of thermal field theory. However, in
the framework of DSE, people may encounter problems, even though one
can take the CJT effective potential as an approximation.
%
%
In Ref.~\cite{YCL06}, some of us have shown that the chiral
susceptibility could be a criterion to judge whether the chiral
phase transition takes place or not. In this section, after
studying the property of the thermodynamic potential in the NJL
model, we would see how much chiral susceptibility could do as the
criterion in judging the phase transition.
%

For a system with volume $V$, temperature $T$ and chemical potential
$\mu$, we could define the thermodynamic potential as
\begin{equation}
\Omega(T,\mu)=-\frac{T}{V} ln \textbf{Tr} \big[
\exp(-\frac{1}{T}\int d^3 x(\mathscr{H}-\mu q^\dag q))\big] \, ,
\nonumber
\end{equation}
where $\mathscr{H}$ is the Hamiltonian corresponding to the
Lagrangian $\mathscr{L}$.

In the conventional way, as we take only the scalar condensate into
account in the NJL model, we have
\begin{equation}
\mathscr{L}+\mu q^\dag q=\bar{q}(i\gamma^\mu \partial_\mu-M)q+\mu
q^\dag q-G\phi^2\nonumber
\end{equation}
where  $\phi=\langle\bar{q}q\rangle, M=m-2G\phi$. The thermodynamic
potential could then be written as
\begin{equation}
\Omega(T,\mu;M)=\Omega_M(T,\mu)+\frac{(M-m)^2}{4G}
\end{equation}
in which
\begin{eqnarray}
\Omega_M(T,\mu)&=&-2 N_c N_f \int \frac{d^3p}{(2\pi)^3}\left\{E_p
            +T ln[1 + \exp(-\frac{1}{T}(E_p-\mu))] \right. \nonumber \\
        &&{}+T ln[1 + \exp(-\frac{1}{T}(E_p+\mu))]\left. \right\} \, .
\end{eqnarray}

Recalling the gap equation discussed in last section, at the limit
$T=0$, we have
\begin{equation}
\Omega(T,\mu,\phi)=\frac{(M-m)^2}{2G}-8 N_c \int^\Lambda_0 \frac{p^2
dp}{2\pi^2}E_p \, , \qquad \qquad \mbox{for} \; \mu<|M|,
\end{equation}
\begin{equation}
\Omega(T,\mu,\phi)=\frac{(M-m)^2}{2G}-8 N_c \int^\Lambda_{k_f}
\frac{p^2 dp}{2\pi^2}E_p-8 N_c \mu \int^{k_f}_0 \frac{k^2
dk}{2\pi^2}\, , \qquad \mbox{for} \; \mu > |M| \, .
\end{equation}

It has been well known that, the stable physics state is the one
corresponds to the global minimum of the thermodynamic potential,
i.e., the one built upon the smallest $\Omega$ determined by the
conditions $\frac{\partial{\Omega}}{\partial M} = 0$  and
$\frac{\partial^{2}{\Omega}}{\partial M^{2}} > 0$. It is easy to
show that the gap equations expressed in Eq.~(6), Eq.~(7) is just
the result of $\frac{\partial{\Omega}}{\partial M} = 0$ for the
corresponding $\Omega$. Then, from the thermodynamic potential, we
could easily compare the solutions, and find out the stable one. In
turn, we could study the phase transition by discussing the
variation of the gap-equation's solutions with respect to the
chemical potential. For the above reasons, we analyze the
characteristic of the thermodynamic potential $\Omega$ against the
constituent quark mass $M$ in the follows.

\subsection{In the Chiral Limit}

From Eqs.~(11) and (12), we could give the thermodynamic potential
in the chiral limit in NJL model just by setting $m=0$. The obtained
thermodynamic potential $\Omega$ in term of the constituent quark
mass $M$ at several values of the chemical potential is illustrated
in Fig.~5 (To guide the eye and for the convenience of discussion,
we have shifted the curves vertically to get zero when $M=0$). From
Fig.~5, we could see that, as the chemical potential is smaller than
$335.59$~MeV, the Wigner solution corresponds always to the maximum
of the thermodynamic potential. With the increasing of $\mu$, two
convexities are separated from the Wigner solution and left a local
minimum there. It means that, as $\mu$ is not large enough, the
Wigner solution is only a metastable state, and the Nambu solutions
are the stable one since it corresponds to the global minimum of the
thermodynamic potential. As we increase the chemical potential
continually, the well generated upon the minimum corresponding to
the Wigner solution becomes deeper and that corresponding to the
Nambu solution gets shadower. As $\mu = 368.00$~MeV, the three
minima of the thermodynamic potential take the same value. When
$\mu$ reaches $397.23$~MeV, the well built upon the Nambu solution
disappears. Accordingly, that of the Wigner solution becomes the
global minimum.
\begin{figure}[!ht]
\includegraphics[scale=1.0,clip]{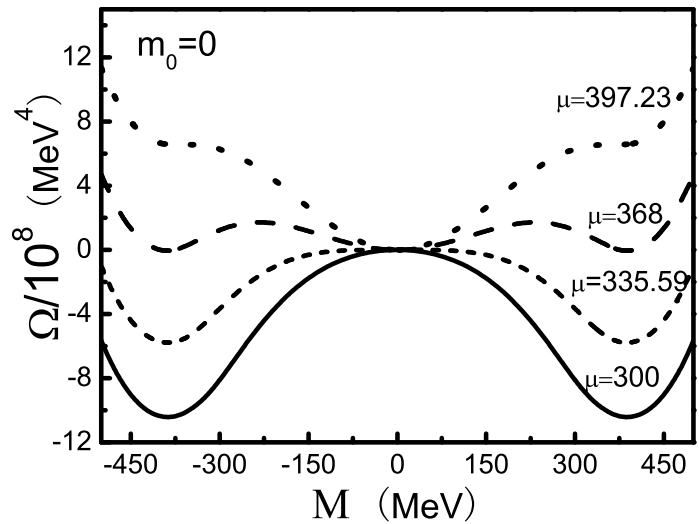}
\caption{Constituent quark mass dependence of the thermodynamic
potential in the chiral limit at several chemical potential (with
the same interaction parameters as those for Fig.~1).}
\end{figure}
Such a characteristic of the evolution of the thermodynamic
potential in terms of the chemical potential and the constituent
quark mass indicates that, at small chemical potential, the system
is in the chiral symmetry broken phase since the Nambu solution
corresponds to the stable state. As the chemical potential is larger
than a critical value, the system appears in the chiral symmetry
restored phase since the Wigner solution corresponds to the stable
state. And the chiral phase transition is in first order.

Comparing Figs.~1, 2 and 5, we could infer that the chiral
susceptibility $\chi$ could describe the local property of the
relation between the thermodynamic potential $\Omega$ and the
constituent quark mass $M$. At Wigner solution, the divergence of
$\chi$ from negative to positive represents the process for the
thermodynamic potential to change from convex one to concave one.
And it shows that the state with $M=0$ changes from a unstable state
to a metastable state after the chemical potential raises to
$335.59$~MeV. We can recognize then that the chiral susceptibility
could present the possibility for the phase transition to happen but
the chemical potential for the $\chi$ to be divergent is not the
critical one for the phase transition to take place. In this sense,
the $\chi$ is a quantity to describe the local property of the gap
equation's solutions. In realistic physics problem, the system may
not transfer to the global minimum when phase transition happens,
but may stay at a metastable state for a while. Thus we could take
the chiral susceptibility to study the existence of the possible
states, and from this property, we could generally get the
information of the relation between the thermodynamic potential
$\Omega$ and the constituent quark mass $M$.

\subsection{Beyond the Chiral Limit}

The similar thing happens in the situation beyond chiral limit. With
Eqs.~(11) and (12) and the parameters used in last section, we
evaluate the thermodynamic potential as a function of the
constituent quark mass at various chemical potentials in the case of
beyond the chiral limit in the NJL model. The obtained results are
illustrated in Fig.~6.
\begin{figure}[!ht]
\includegraphics[scale=1.0,clip]{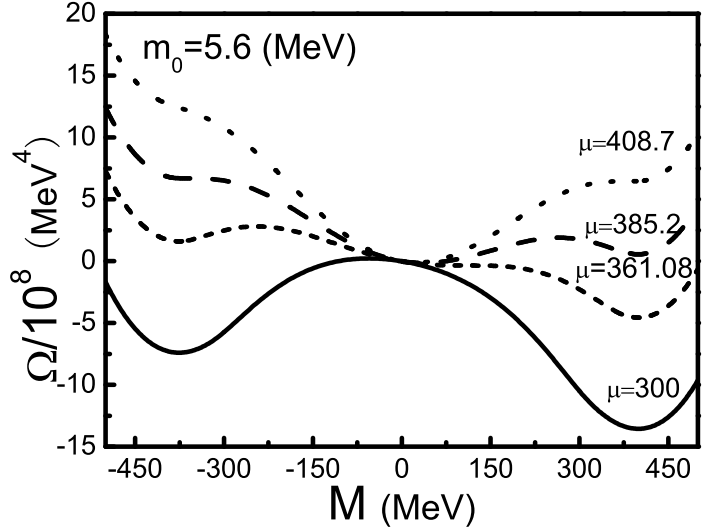}
\caption{Constituent quark mass dependence of the thermodynamic
potential beyond the chiral limit at several chemical potentials
(with the same parameters as those for Fig.~2 and a current quark
mass $m=5.6$~MeV. }
\end{figure}
It is evident that the general feature of the thermodynamic
potential beyond the chiral limit is similar to that in the chiral
limit (shown in Fig.~5), however it is a little bit more complicated
in details. When the chemical potential is small, the thermodynamic
potential holds three extrema, two of them are in the concave and
corresponds to the positive, negative Nambu solution, respectively,
the other is in the convex and corresponds to the Wigner solution.
It indicates apparently that the Nambu solutions are stable states
and the Wigner solution is a unstable state. When $\mu$ takes a
value $361.08$~MeV, a inflexion emerges at $M=112.39$~MeV which
separates the curve of the thermodynamic potential into convex part
and concavity part. Recalling the solutions of the gap equation
shown in Fig.~3, we know that such a point corresponds to the one
for a bifurcation to appear. However, as the chemical potential is
less than $382$~MeV, the extremum corresponding to the positive
Nambu solution is always the global minimum. $\mu = 382$~MeV is thus
the critical value for the chiral phase transition to takes place,
since the concave generated from $\mu=361.08$~MeV becomes the global
minimum from then on. When the chemical potential reaches
$385.20$~MeV and $408.70$~MeV, the concavities corresponding to the
negative, the positive Nambu solution merges with convexities
respectively. Thus the unstable-states vanish at the corresponding
point, respectively.

Combining Fig.~6 with Fig.~4, we could recognize that the chiral
susceptibility can describe the possibility of the state
generation and disappearance well. In detail, the singularity of
the $\chi$ corresponds to such generation and annihilation but not
the critical point for the chiral phase transition to happen. It
is obvious that such a result is as the same as that in the chiral
limit.

\section{Relation Between the Chiral Susceptibility and the Thermodynamic Potential}

In the last two sections, we have discussed the solutions of the gap
equation, the thermodynamic potential and the chiral susceptibility
in the NJL model, and shown that the chiral susceptibility could be
used to study the property of gap equation numerically. It is
certain that it would be helpful to understand the significance of
the chiral susceptibility, if we could find an analytical relation
between the chiral susceptibility and the thermodynamic potential.
Then we do this in this section.

It has been shown that Eq.~(11) and Eq.~(12) represent the
explicit form of thermodynamic potential without any
approximations in the NJL model. From them, we could get the gap
equation by evaluating the first order derivative of the $\Omega$
over the $M$ and setting it to zero. It reads explicitly as
\begin{eqnarray}
\frac{\partial \Omega}{\partial M}& =& \frac{M-m}{G}-4 N_c N_f
\int \frac{d^3p}{(2\pi)^3}\frac{M}{E_p}[1-n_p(T,\mu)
                -\bar{n_p}(T,\mu)] \nonumber \\
                                 & \equiv  & 0  \, .
\end{eqnarray}
Furthermore, we get the explicit form of the second order
derivative $\frac{\partial^2 \Omega}{\partial M^2}$ as
\begin{eqnarray}
\frac{\partial^2 \Omega}{\partial M^2}& =& \frac{1}{G}-4 N_c N_f
\frac{\partial}{\partial M}\Big\{ \int\frac{d^3p}{(2\pi)^3}
\frac{M}{E_p}[1-n_p(T,\mu)-\bar{n_p}(T,\mu)] \Big\} \, .
\end{eqnarray}

From Eq.~(13) and the definition of chiral susceptibility, we obtain
the expression of $\chi$ by doing the first order derivative of
Eq.~(13) over $m$ as
\begin{eqnarray}
\chi&=&\frac{\partial M}{\partial m} = \frac{1}{1 - 4 G N_c N_f
\frac{\partial}{\partial M} \Big\{
\int\frac{d^3p}{(2\pi)^3}\frac{M}{E_p}[1-n_p(T,\mu)-\bar{n_p}(T,\mu)]
\Big\} }
 \, .
\end{eqnarray}

From Eq.~(15) and Eq.~(14), we can easily find that chiral
susceptibility has a simple relation with $\frac{\partial^2
\Omega}{\partial M^2}$, which reads
\begin{equation}
\chi=\frac{1}{G \frac{\partial^2 \Omega}{\partial M^2}} \, .
\end{equation}

From the basic mathematical knowledge, we know that the second
order derivative of a function determines the concavo-convexity of
the function. And as a convexity and a concavity merge together,
there will be a bifurcation, which corresponds to an inflexion. If
there are two convexities and one concavity merge at the same
time, there would be an trifurcation. In these two situations,
both of the $\frac{\partial^2 \Omega}{\partial M^2}$ are zero,
thus the chiral susceptibility has a singularity at such kind
point. On the other hand, we know that a local minimal point
corresponds to a point where $\frac{\partial \Omega}{\partial
M}=0$ and $\frac{\partial^2 \Omega}{\partial M^2}>0$. We could
then take the chiral susceptibility as a signature to judge the
the stability of the state in the NJL model.

In the above discussion, we have not take any approximation, for
example, the chiral limit and zero temperature limit, in getting the
thermodynamic potential and the chiral susceptibility. The obtained
relation between the chiral susceptibility and the thermodynamic
potential is general in the NJL model. Even though it seems to be
difficult to obtain such kind explicit relation in the framework of
Dyson-Schwinger equation, where we have difficulty in getting the
explicit form of the thermodynamic potential, the chiral
susceptibility is still useful to determine the generation and/or
annihilation of the possible state.

\section{Summary }

In summary, we have evaluated the solutions of the gap equation, the
thermodynamic potential and the chiral susceptibility in and beyond
the chiral limit at finite chemical potential in the NJL model,
which takes the point-like interaction to approximate the
gluon-mediated interaction among quarks. The evolution, especially
the furcation of the solutions and the divergence of the chiral
susceptibility with respect to the chemical potential are discussed.
It shows that as a furcation emerges to the solutions of the gap
equation, the concavo-convexity of the thermodynamic potential would
change and a divergence would appear in the chiral susceptibility.
Furthermore, we give an explicit relation between the chiral
susceptibility and the thermodynamic potential in the NJL model. It
indicates that the singularity of the chiral susceptibility
corresponds to the point for the furcation of the solutions to
appear as well as for the concavo-convexity of the thermodynamic
potential to change. We could then believe convincingly that the
chiral susceptibility $\chi$ is a quantity being able to
characterize the stability of the possible states, and to judge
which state is the physically allowed one. Furthermore, chiral
susceptibility is a quantity could be generalized to other
frameworks, such as the Dyson-Schwinger equation approach, even
though it is difficult to write down the effective potential
explicitly.

\bigskip

This work was supported by the National Natural Science Foundation
of China under the contract Nos. 10425521 and 10575004, the Major
State Basic Research Development Program under contract No.
G2000077400, the Key Grant Project of Chinese Ministry of Education
(CMOE) under contact No. 305001 and the Research Fund for the
Doctoral Program of Higher Education of China under grant No.
20040001010. One of the authors (Y. Zhao) acknowledges also the
support by Peking University and Mr. Qi-min Quan for his helpful
discussions. One of the authors (Y.X. Liu) thanks the support by the
CMOE, too.


\end{document}